\pdfoutput=1

\newif\ifieeestyle
\ieeestyletrue 
\ifieeestyle
	\documentclass[journal]{IEEEtran}
\else
	\documentclass[english]{article}
\fi

\usepackage{cite}

\usepackage[pdftex]{graphicx}
\DeclareGraphicsExtensions{.pdf,.jpeg,.png}

\usepackage[cmex10]{amsmath}
\usepackage{url}
\usepackage[hidelinks]{hyperref}  
\hypersetup{
pdfauthor={Tilman Johannes Sumpf},
pdftitle={Fast T2 Mapping with Improved Accuracy Using Undersampled Spin-echo MRI and Model-based Reconstructions with a Generating Function},
pdfkeywords={FSE, indirect echoes, model-based reconstruction, relaxometry, T2 mapping}}

\usepackage{siunitx}  
\usepackage{textcomp} 
\usepackage{multirow} 

\newif\ifmathvars
\ifmathvars
	\newcommand{\Bo}{\ensuremath{\mbox{B}_1} }
	\newcommand{\Bop}{\ensuremath{\mbox{B}_{1,\op{peak}}} }
	\newcommand{\To}{\ensuremath{\mbox{T}_1} }
	\newcommand{\Tt}{\ensuremath{\mbox{T}_2} }
	\newcommand{\Ltwo}{\ensuremath{L^2} }	
	\newcommand{\z}{\ensuremath{z}}			
	\newcommand{\kspace}{\ensuremath{k\mbox{-space}}}
\else
	\newcommand{\Bo}{\ensuremath{\mbox{B1}} }
	\newcommand{\Bop}{\ensuremath{\mbox{B1}_{\op{peak}}}}
	\newcommand{\To}{\ensuremath{\mbox{T1}} }
	\newcommand{\Tt}{\ensuremath{\mbox{T2}} }
	\newcommand{\Ltwo}{\ensuremath{\mbox{L2}} }
	\newcommand{\z}{\ensuremath{\mbox{z}}}
	\newcommand{\kspace}{\ensuremath{\mbox{k-space}}}
\fi

\newcommand{\refeq}[1]{(\ref{#1})} 
\newcommand{\mb}{\mathbf}
\newcommand{\mbs}{\boldsymbol}
\newcommand{\op}{\operatorname}

\begin{document}

\title{Fast \Tt Mapping with Improved Accuracy Using Undersampled Spin-echo MRI and Model-based Reconstructions with a Generating Function}

\author{Tilman~J.~Sumpf*,
        Andreas~Petrovic,
       	Martin~Uecker,
       	Florian~Knoll,
        and~Jens~Frahm
\thanks{Manuscript submitted May 13, 2014. \it{Asterisk indicates corresponding author.}}
\thanks{*T.J. Sumpf (e-mail: \url{tsumpf@gwdg.de}) and J. Frahm are with the Biomedizinische NMR Forschungs GmbH, Max-Planck-Institute for Biophysical Chemistry, Am Fassberg 11, 37077 G\"ottingen, Germany.}        
\thanks{A. Petrovic and F. Knoll are with the Institute for Medical Engineering, Graz University of Technology, Austria. A. Petrovic is also with the Ludwig Boltzmann Institute for Clinical Forensic Imaging, Graz, Austria. F. Knoll is also with the Center for Biomedical Imaging, New York University School of Medicine, New York.}
\thanks{M. Uecker is with the Department of Electrical Engineering and Computer Sciences, University of California, Berkeley, California.}}

\maketitle

\begin{abstract}
A model-based reconstruction technique for accelerated \Tt mapping with improved accuracy is proposed using undersampled Cartesian spin-echo MRI data. The technique employs an advanced signal model for \Tt relaxation that accounts for contributions from indirect echoes in a train of multiple spin echoes. An iterative solution of the nonlinear inverse reconstruction problem directly estimates spin-density and \Tt maps from undersampled raw data. The algorithm is validated for simulated data as well as phantom and human brain MRI at 3\,T. The performance of the advanced model is compared to conventional pixel-based fitting of echo-time images from fully sampled data. The proposed method yields more accurate \Tt values than the mono-exponential model and allows for undersampling factors of at least 6. Although limitations are observed for very long \Tt relaxation times, respective reconstruction problems may be overcome by a gradient dampening approach. The analytical gradient of the utilized cost function is included as Appendix.
\end{abstract}

\ifieeestyle
\fi

\section{Introduction}
\ifieeestyle
	\IEEEPARstart{Q}{uantitative}
\else
	Quantitative
\fi
evaluations of \Tt relaxation times are of importance for an increasing number of clinical MRI studies \cite{cheng}. Conventional \Tt mapping relies on the time-demanding acquisition of fully sampled multi-echo spin-echo (MSE) MRI datasets. Recent advances exploit a model-based reconstruction strategy which allows for accelerated \Tt mapping from undersampled data \cite{block, doneva, sumpf:martini, huang:pca}. So far, however, these approaches have been limited to a mono-exponential signal model, whereas experimental spin-echo trains are well known to deviate from the idealized behavior, for example, because of \Bo inhomogeneities and non-rectangular slice profiles \cite{majumdar:mserr1, majumdar:mserr2, crawley}. Under these circumstances, model-based reconstructions lead to \Tt errors even for fully sampled conditions and additional deviations for different degrees of undersampling. As a consequence, it is common practice to discard the most prominently affected first echo of a MSE acquisition \cite{sumpf:martini, smith, mosher}.
 
To overcome the aforementioned problem, an analytical formula has been proposed which models the MSE signal more accurately \cite{lukzen,petrovic}. This work demonstrates the application of this advanced model \cite{petrovic} for a model-based reconstruction. The new approach allows for highly accelerated as well as accurate \Tt mapping from undersampled Cartesian MSE MRI data. Part of this work has been presented in \cite{sumpf:gf}. A related approach was recently proposed in \cite{huang:phasegraph}, where the extended phase graph (EPG) model is used in a dictionary-based linear reconstruction algorithm. First promising results for a nonlinear inversion of the computationally demanding EPG algorithm have been reported in \cite{lankford} for small matrices ($64 \times 64$ samples).
\begin{figure}[!t]
\centering
\includegraphics[width=\columnwidth]{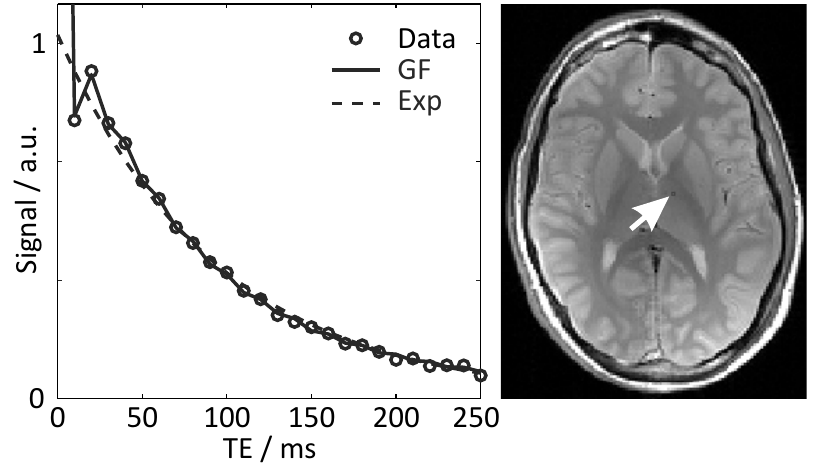}
\caption{Generating function (GF) and single exponential (Exp) fitted to the magnitude signal (circles) of a multi-echo spin-echo MRI acquisition of the human brain (25 echoes, single pixel = arrow). Because the GF is only valid at exact echo times, the solid curve is an interpolation for display purposes.}
\label{fig1}
\end{figure}

\section{Theory}
\Tt mapping relies on a train of successively refocused spin echoes. In practical MRI, however, the assumption that these signals represent the true \Tt relaxation decay is violated due to the formation of indirect echoes \cite{henning:1,henning:2}. The most notable consequence is a hypointense first spin echo which clearly disrupts the expected mono-exponential signal decay. A variety of attempts to better reproduce the echo amplitudes employ the EPG algorithm \cite{woessner,kiselev,zur,lebel} which considers different magnetization pathways in a recursive description. In 2007 Lukzen et al. \cite{lukzen} obtained an explicit analytical expression for the problem by exploiting the generating function (GF) formalism \cite{comtet,wilf}. In contrast to the EPG algorithm, the model formula can be implemented very efficiently using the fast Fourier transform. An example of the improved performance for an extended GF model \cite{petrovic} which includes non-ideal slice profiles is shown in Fig.~\ref{fig1} for human brain MRI data.

\subsection{Generating function}
The GF for the MSE signal amplitudes is given in the \z -transform domain \cite{lukzen}
\begin{equation}
\label{eq:gf:base}
\begin{split}
& G_{\rho,\alpha,k_1,k_2 } (z)=
\frac{\rho}{2} \\
&+\frac{\rho}{2}
\sqrt{
\frac{(1+zk_2) \left[ 1-z(k_1+ k_2 ) \cos \alpha+ z^2 k_1 k_2 \right]}
{(-1+z k_2 )   \left[ -1+z(k_1 - k_2 ) \cos \alpha +z^2 k_1 k_2\right]}
} 
\mbox{,}
\end{split}
\end{equation}
where $\rho$ is the spin density, $k_1 = \exp(-\tau/\To)$ and $k_2 = \exp(-\tau/\Tt)$ are the \To and \Tt relaxation terms, $\alpha$ is the refocusing flip angle and $\tau$ the echo spacing. $z$ denotes a complex variable in the $z$-domain. Evaluation of \refeq{eq:gf:base} on the unit circle, i.e. for $z = \exp(i \omega)$, $\omega= 0 \hdots 2 \pi$, yields a frequency-domain representation of the echo amplitudes at echo times TE. 

The non-uniform flip-angle distribution of a non-ideal slice profile can be accounted for by superimposing evaluations of \refeq{eq:gf:base} for different values of $\alpha$ \cite{petrovic}. The final formulation in frequency domain is given by
\begin{equation}
\label{eq:gf:profile}
S_{\rho,k_2 } (\omega)= \frac{1}{Q}  \sum_{q=1}^Q G_{\rho,\alpha_q,k_1,k_2 } (e^{i\omega })\mbox{,}
\end{equation}
with $\alpha_q$ a finite number of $Q$ supporting points characterizing the profile of the refocusing pulse in slice direction. The values for $k_1$ and the absolute magnitudes of the angles $\alpha_q$ have to be taken from experimentally determined \To and \Bo maps. The echo amplitudes in time domain can be recovered by application of a discrete Fourier transform (DFT) on the resulting frequency-domain samples.
 
Given a series of magnitude images from a MSE train, the GF approach can be used to determine quantitative \Tt values at different spatial positions by pixel-wise fitting. The method has been demonstrated to yield more accurate \Tt estimations than a mono-exponential fit \cite{petrovic}. As a potential limitation, \refeq{eq:gf:profile} requires a valid \To and \Bo map prior to \Tt reconstruction as well as an estimation of the pulse profile in slice direction. The influence of errors in these estimates on the reconstructed \Tt maps has previously been elaborated for fully sampled data \cite{petrovic}.

\subsection{Reconstruction from undersampled data}
In addition to the DFT $F_\omega$ along the samples in frequency domain, \refeq{eq:gf:profile} can be extended by a two-dimensional DFT $F_{xy}$ to synthesize \kspace samples from estimated parameter maps. Similar to the approaches described in \cite{block,sumpf:martini} the conformity of these (synthetic) data with the experimentally available samples $\mb{s_c}$ from a MSE acquisition can be quantified with a cost function
\newcommand*{\bx}{\mb{x}}
\begin{align}
\Phi(\bx) &= \frac{1}{2}\sum_c \| M(\bx)-\mb{s_c} \|_2^2 \label{eq:cost}\\
M(\bx)    &= P F_{xy} C_c F_{\omega} W(\bx) \\
\bx       &= \left(\begin{array}{cc} \mbs{\rho}\\\mb{k_2}\\ \end{array} \right) \label{eq:x}\\
\mbs \rho &= \left( \rho_1, \hdots , \rho_{N_p} \right)\\
\mb{k_2}  &= \left( k_{2,1}, \hdots , k_{2, N_p} \right)^T
\end{align}
where the diagonal matrices $P$ and $C_c$ contain the binary sampling pattern and the complex coil sensitivities of the coil elements $c$, respectively. $W(\bx)$ represents the combined results from evaluating \refeq{eq:gf:profile} at all $N_p$ pixel positions and $N_\omega$ frequencies. Minimization of \refeq{eq:cost} with respect to the components of $\mb x$ allows for the direct reconstruction of $\Tt  =-\tau / \op{ln}⁡(k_2)$ and $\rho$ parameter maps from undersampled data.

\subsection{Column-wise reconstruction} 
When using Cartesian sampling schemes where undersampling is only performed in the phase-direction ($y$), the Fourier transform $F_x$ in read-direction can be excluded from the cost function \refeq{eq:cost} and replaced by a respective inverse DFT of the data samples $\mb{s_c}$ prior the iterative reconstruction. This approach not only reduces the computational costs for the evaluation of each cost function, but also allows for an independent and parallel reconstruction of image columns. This strategy splits the overall image-reconstruction into much smaller problems, which usually converge significantly faster than the respective global optimization approach. Another advantage is the possibility to remove noise columns from the reconstruction, e.g. by masking columns with an overall energy below a given threshold. 

\subsection{Oversampling on the \z -plane}
As stated before, evaluation of the GF on the unit circle in \z -domain allows for the calculation of MSE amplitudes by application of a DFT in \z -direction. The range of echo times is inversely proportional to the frequency resolution, so that for $N_\omega$ frequency samples the longest modeled echo time yields
\newcommand{\TEmax}{\mbox{TE}_{\op{max}}}
\begin{equation}
\TEmax = N_\omega \tau \mbox{.}
\end{equation}
As a rule of thumb, $\TEmax$  should be at least $6$ times the longest \Tt within the measured object to ensure a proper coverage of the \Tt signal decay. If this requirement is violated, the modeled echo amplitudes (i.e., the DFT of the GF) become distorted due to aliasing in time direction. In practice, typical \Tt mapping protocols use $16$ echoes with an echo spacing of $\SI{10}{ms}$. To avoid reconstruction errors for \Tt values longer than about $\SI{30}{ms}$, it is therefore reasonable to evaluate \refeq{eq:gf:base} for a considerably higher number of frequency samples than the number NE of actually measured echoes. Unfortunately, this oversampling on the \z -plane yields a substantial increase of the computational costs. The current implementation therefore applied only a moderate oversampling ($N_\omega = 128$) to define \Tt values of up to $\SI{213}{ms}$ ($\tau = \SI{10}{ms}$) which cover most tissues in brain \cite{poon} and other organ systems except for fluid compartments. In fact, the precise determination of these very long \Tt values has only limited clinical relevance. For conventional pixel-wise fitting it is therefore reasonable to accept quantitative errors in regions with \Tt values above this limit. For model-based reconstructions, however, pixels with an implausible signal behavior or long \Tt cannot simply be excluded from the vector of unknowns. Respective artifacts, in particular for reconstructions from undersampled data, need to be treated by additional means (see below).

\section{Materials and Methods}
\subsection{Optimization and gradient scaling}
In accordance to \cite{block, sumpf:martini} we used the CG-DESCENT algorithm \cite{hager} to minimize the nonlinear cost function \refeq{eq:cost}. This method offers a guaranteed descent and an efficient line search but requires the gradient of the objective function to be balanced with respect to its partial derivatives. Therefore, the vectors $\mb{k_2}$ and $\mbs{\rho}$ in \refeq{eq:x} have been substituted by scaled variants
\begin{align}
\mbs{\tilde{\rho}} &= L_{\rho}^{-1} \mbs\rho\\
\mbs{\tilde{k}_2} &= L_{k}^{-1} \mb{k_2}
\end{align}
with $L_\rho$ and $L_k$ being diagonal scaling matrices. The dimensioning of the scaling involves several challenges. For example, we found large \Tt values ($\Tt > \TEmax /6$) to provoke very high entries in the cost function gradient, so that even few discrete regions may lead to a global failure of the reconstruction process when using scalar values for the scaling. To counterbalance these effects, a dynamic validity mask has been implemented to detect potentially destructive pixels during reconstruction and to reduce the gradient scaling in respective regions. Because such regions are initially unknown for undersampled data, the reconstruction was performed in three steps with a fixed number of $3 \times 20$ CG-iterations for each image column. After each of the three optimization blocks, the validity mask was updated and the gradient scaling was dampened by a factor of $10^{-5}$ in regions with a \Tt of either more than $\TEmax /6$ or less than $\SI{0}{ms}$. With all data initially standardized by their overall \Ltwo norm, the scaling was initialized with heuristically chosen scalar values of $L_\rho = 1$  and $L_k = 40$.

Pixel-wise fitting of echo-time images involved the Matlab nlinfit program (MathWorks, Natick, MA), i.e. the Levenberg Marquard algorithm. In contrast to the CG-DESCENT optimization approach, the gradient was approximated using the internal finite-difference method of nlinfit rather than the explicit analytical expression given in the Appendix.

\subsection{Regularization}
\renewcommand*{\l}{\lambda_}
Especially in the first optimization block, the model insufficiencies for large \Tt values can lead to strong outliers in the reconstructed maps. The effect can be reduced by penalizing the \Ltwo norm of the difference between estimates $\mbs\rho$ and $\mb{k_2}$ and their initial guesses $\mbs{\rho_0}$ and $\mb{k_{2,0}}$. The technique, known as Tikhonov regularization, introduces the regularization parameters $\l \rho$ and $\l k$ to the cost function:
\begin{equation}
\begin{split}
\Phi(\bx) &= \frac{1}{2}\sum_c \| M(\bx)-\mb{s_c} \|_2^2 \\
&+\l\rho \|\mbs{\rho} - \mbs{\rho_0} \|_2^2 + \l k \| \mb{k_2} - \mb{k_{2,0}} \|_2^2 \label{eq:costreg}
\end{split}
\end{equation}
The parameters balance the data fitting term and the penalty terms and need to be tuned accordingly. Here, we chose an iteratively regularized approach with starting values of $\l\rho= 1 \cdot 10^{-3}$ and $\l k = 3 \cdot 10^{-3}$. After each of the three optimization blocks, both scaling factors are reduced by a factor of $10^{-3}$. This yields a moderate regularization at the beginning (far from the solution) and only minimal regularization at the end.

\subsection{Numerical simulations}
To test the algorithm against different conditions, we used simulated data for a numerical phantom with multiple objects exhibiting equal spin density but different \Tt relaxation times ranging from $50$ to $\SI{800}{ms}$. Simulated noiseless \kspace samples for a $160 \times 160$ data matrix were derived from superimposed circles using the analytical Fourier space representation of an ellipse \cite{slaney,walle}. The signal amplitudes in time domain were derived using \refeq{eq:gf:profile} with subsequent DFT as a forward model. To avoid aliasing in time direction, the simulated data were derived from $4096$ frequency-domain samples for every phantom compartment. Only the first $17$ points were further used in time domain, corresponding to echo times of $10$ to $\SI{170}{ms}$. The simulated profile for the refocusing pulse was derived from a Gaussian using $16$ support points. The \To map was simulated to yield a constant value of $\To = \SI{1000}{ms}$ throughout the FOV. The \Bo field was simulated to be homogeneous and to correspond to an ideal flip angle of $\SI{180}{\degree}$ at the center of the slice profile. 

\subsection{MRI measurements}
To experimentally validate the accuracy of the approach using multi-channel data, we used MRI data of a commercially available relaxation phantom (DiagnosticSonar, Eurospin II, gel nos $3$, $4$, $7$, $10$, $14$ and $16$). The phantom contains $6$ compartments with \Tt values ranging from $46$ to $\SI{166}{ms}$ and \To values ranging from $311$ to $\SI{1408}{ms}$ (at $21$\si{\degree}C). For human brain MRI, a young healthy volunteer with no known abnormalities participated in this study and gave written informed consent before the examination.

All MRI experiments were conducted at $\SI{3}{T}$ (Tim Trio, Siemens Healthcare, Erlangen, Germany). While radiofrequency excitation was accomplished with the use of a body coil, signal reception was performed by a $12$-channel head coil in CP mode (circular polarized), thus yielding data from 4 virtual elements. The field map was acquired using the method by Bloch-Siegert \cite{sacolick} (Gaussian pulse, \Bop $= \SI{0.11}{G}$, $K_{\op{BS}} = \SI{21.3}{rad G^{-2} ms^{-1}}$, $\SI{8}{s}$ duration, $f_{\op{OR}} = \SI{8}{kHz}$ , FOV = $250 \times \SI{250}{mm^2}$, matrix $= 64 \times 32$, slice thickness $= \SI{8}{mm}$, $\alpha = \SI{60}{\degree} / \SI{120}{\degree}$, scan time $= \SI{13}{s}$). \To values were obtained using a Turbo Inversion Recovery (TIR) sequence (TR $= \SI{7}{s}$, TE $= \SI{7.8}{ms}$, TI $= 100-\SI{3100}{ms}$, turbo factor $= 5$, FOV $= 250 \times \SI{250}{mm^2}$, matrix $= 192 \times 192$, slice thickness $= \SI{4}{mm}$, scan time $= 17\mbox{:}\SI{05}{min}$) and a traditional magnitude fitting procedure. The data samples $s_c$ were acquired using a MSE sequence (TR $= \SI{4}{s}$, echo spacing $= \SI{10}{ms}$, $25$ echoes, FOV $= 250 \times \SI{250}{mm^2}$, matrix $= 192 \times 192$, slice thickness $= \SI{4}{mm}$, scan time $= 12\mbox{:}\SI{54}{min}$). For the relaxation phantom, $16$ additional (single) spin-echo (SE) experiments were conducted with the same parameters and equally spaced echo times TE from $10$ to $\SI{160}{ms}$ (scan time $= 206\mbox{:}\SI{24}{min}$).

\subsection{Reconstruction and undersampling}
\begin{figure}[!t]
\centering
\includegraphics[width=\columnwidth]{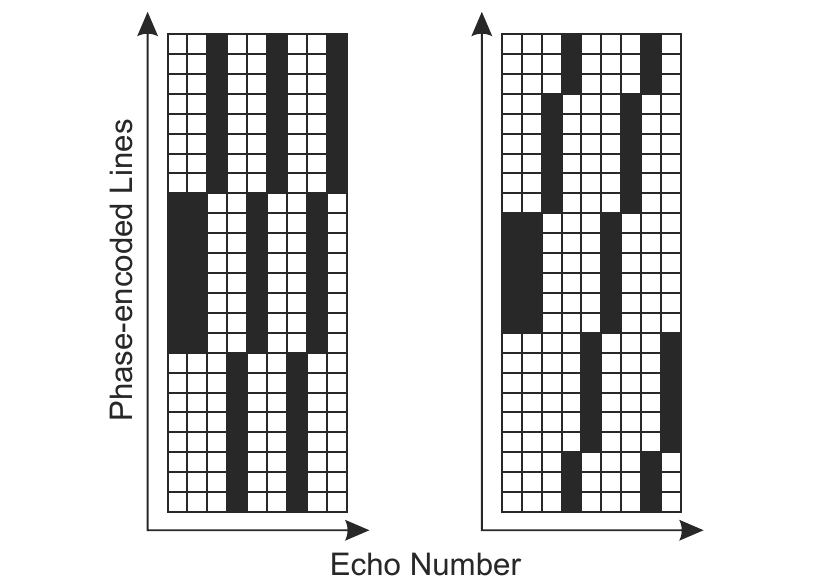}
\caption{Cartesian encoding scheme with a blocked undersampling pattern for acceleration factors of 3 (left) and 4 (right). The examples refer to a \kspace of 24 phase-encoded lines and 9 echoes. Solid symbols represent acquired lines, while open symbols refer to lines not measured.}
\label{fig2}
\end{figure}

Apart from conventional mono-exponential fitting of echo-time images, simulated and measured SE and MSE data were analyzed using the proposed model-based reconstruction with the GF model (GF-MARTINI = Model-based Accelerated Relaxometry by Iterative Nonlinear Inversion). Undersampling for acceleration factors R of up to $12$ was simulated using a "block" pattern as depicted in Fig.~\ref{fig2}.
In contrast to the scheme used for mono-exponential MARTINI \cite{sumpf:martini}, the pattern was designed such that the first two echo times are sampled around the \kspace center. In combination with the GF model, this strategy was found to minimize deviations in the \Tt estimation at different undersampling factors.

The initial spin-density map for GF-MARTINI was set to zero. The map for $k_2$ was initialized with a value that corresponds to $\TEmax /6$ for all pixels. The maps for \To and \Bo were limited to $\SI{100}{ms} <$ \To $< \SI{5000}{ms}$. \Bo values were limited to correspond to refocussing flip angles of at least $\SI{90}{\degree}$ in the center of the assumed Gaussian slice profile. Coil sensitivities were estimated in a pre-processing step according to \cite{uecker} using a fully sampled composite dataset derived from the mean samples of all echoes. The algorithm was constrained to shift all phase information into the complex coil sensitivities, while all parameter maps were assumed to be real. Finally, the sensitivity map of each receiver channel was standardized by the root sum square (RSS) of all channels.

Reference maps from fully sampled data were created by applying the nlinfit tool to magnitude images from the RSS of all receiver channels. For a constant number of echoes in time domain, GF-MARTINI was performed for different numbers of frequency-domain samples with and without the proposed validity mask. For mono-exponential fits the first echo was discarded.

\section{Results}
\subsection{Oversampling and adaptive masking}
\begin{figure*}[bt]
\centering
\ifmathvars
	\includegraphics[width=\textwidth]{fig3_sub}
\else
	\includegraphics[width=\textwidth]{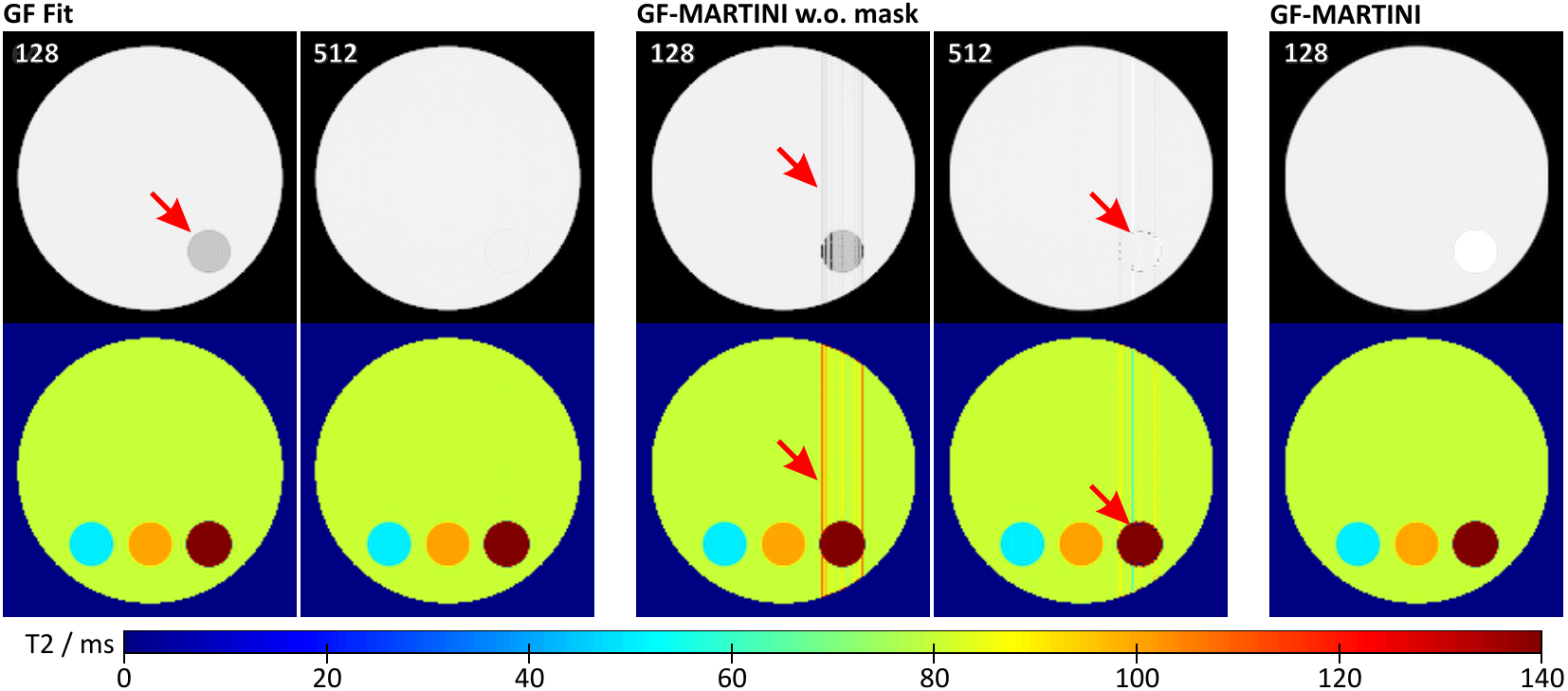}
\fi
\caption{Spin-density and color-coded \Tt maps of a fully sampled numerical phantom with \Tt compartments of 50, 100, and 800\,ms (surrounding: 80\,ms). The maps were obtained by (left) pixel-based GF fitting (128 and 512 frequency-domain samples) as well as GF-MARTINI reconstruction (middle) without and (right) with adaptive validity mask. All artifacts due to model violations can be avoided by the proposed method. For details see text.}
\label{fig3}
\end{figure*}

Figure~\ref{fig3}
shows spin-density and \Tt maps of a noiseless numerical phantom reconstructed by pixel-wise GF fitting and the proposed GF-MARTINI method without and with the adaptive validity mask. The $17$ time-domain data points (echoes) per pixel were derived from either $128$ or $512$ samples in the frequency-domain that define the degree of oversampling on the \z -plane. As expected, even for the pixel-based reconstruction (Fig.~\ref{fig3}, left), both the spin-density and \Tt map yield errors for the compartment with the longest \Tt of $\SI{800}{ms}$ when using only $128$ samples ($128$, arrow). However, \Tt values of the remaining compartments were accurately reconstructed with less than $\SI{2}{ms}$ deviation from the true value. The problem may largely be reduced by extensive oversampling with $512$ frequency-domain samples ($512$) as revealed by a mean \Tt value of $\SI{810}{ms}$ in the rightmost compartment, which is less than $\SI{2}{\%}$ deviation from the true value.

Figure~\ref{fig3} (center) shows the corresponding maps for GF-MARTINI when using $60$ iterations per column and an optimized but constant gradient scaling for every pixel (no masking of invalid regions). The results in ($128$) not only demonstrate the expected deviations for the long-\Tt compartment, but also artifacts (arrows) in the remote part of the columns that comprise the compartment. While the effect is again reduced for the higher number of frequency-domain samples ($512$), residual artifacts originating from model violations at the compartment borders persist. 

Finally, Fig.~\ref{fig3} (right) demonstrates that all artifacts can be avoided by the proposed adaptive mask, even for moderate oversampling of $N_\omega= 128$ samples. In this case the numerical results are in good agreement with the corresponding pixel-based GF fit ($128$) except for the masked-out right compartment. Here, the values for GF-MARTINI are mainly influenced by the applied \Ltwo regularization, which is not included in the pixel-based fit.

\subsection{Accuracy of the GF model}
\begin{figure*}[bt]
\centering
\ifmathvars
	\includegraphics[width=\textwidth]{fig4_sub}
\else
	\includegraphics[width=\textwidth]{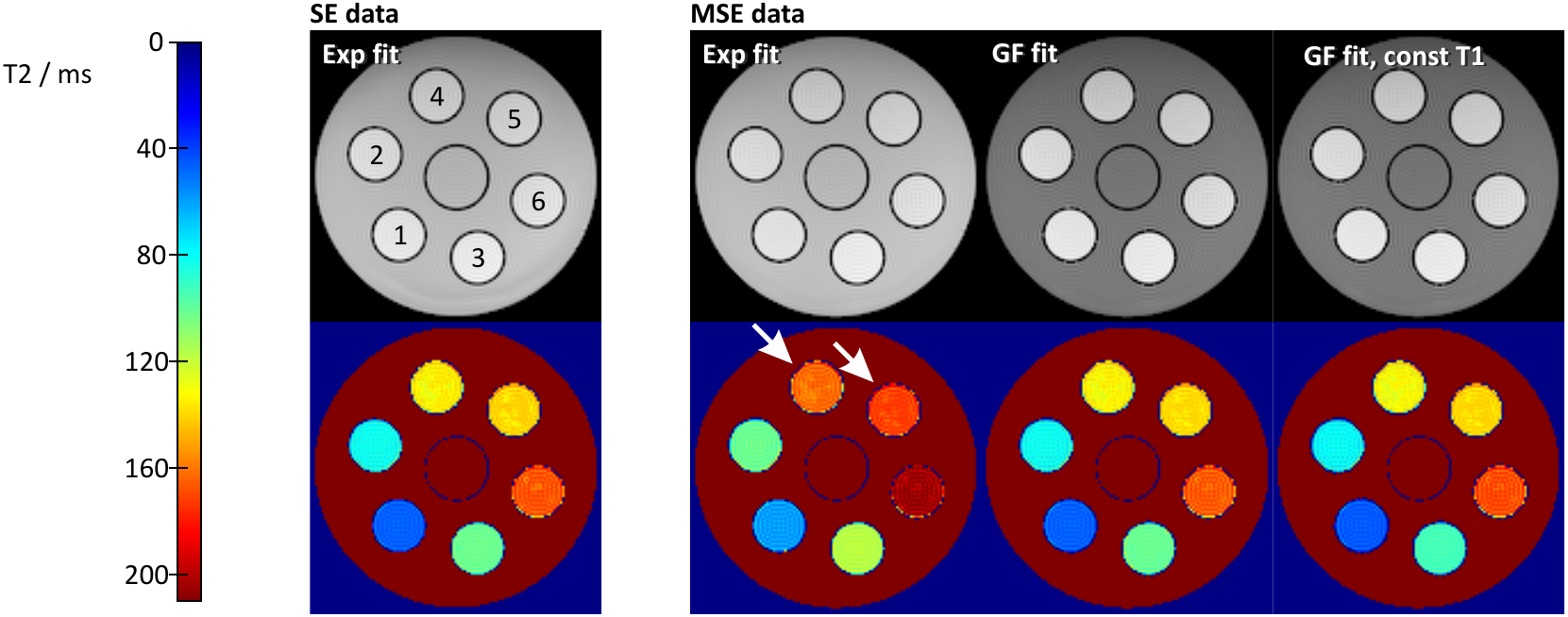}
\fi
\caption{Spin-density and color-coded \Tt maps from fully sampled spin-echo (SE) and multi-echo spin-echo (MSE) data. In comparison to the SE reference, exponential fitting of the MSE data leads to an overestimation of \Tt values (arrows), while GF fitting recovers the true values with high accuracy even when assuming a constant \To = 1000\,ms for all pixels.}
\label{fig4}
\end{figure*}

Figure~\ref{fig4}
compares spin-density and \Tt reconstructions obtained for a fully sampled SE and fully sampled MSE dataset. The results for a pixel-wise mono-exponential fitting of the SE magnitude (measurement time $= \SI{206}{min}$) serve as a reference (ground truth). On the other hand, the corresponding fitting of the MSE images (first echo removed) yields the expected \Tt overestimation in all compartments (arrows). In contrast, the GF fitting of the same data results in \Tt values which are remarkably similar to the reference. Deviations in the surrounding water compartment, which are most notable in the spin-density map, are again an effect of the limited frequency domain oversampling ($N_\omega=128$). The quantitative results from a ROI analysis in Table~\ref{table1}
\begin{table*}[t]
\renewcommand{\arraystretch}{1.3}
\caption{\Tt relaxation times of a phantom for different fitting methods}
\label{table1}
\centering
\begin{tabular}{lcccccc}
\hline
Compartment & 1 & 2 & 3 & 4 & 5 & 6 \\ 
\hline
SE Reference & 46 $\pm$ 2 & 81 $\pm$ 3 & 101 $\pm$ 2 & 132 $\pm$ 5 & 138 $\pm$ 4 & 166 $\pm$ 5 \\[2ex]
\multirow{2}{*}{Mono-Exp} & 59 $\pm$ 4 & 101 $\pm$ 5 & 117 $\pm$ 4 & 161 $\pm$ 5 & 170 $\pm$ 4 & 205 $\pm$ 7 \\ 
 & 28.3\% & 24.7\% & 15.8\% & 22.0\% & 23.2\% & 23.5\% \\ [2ex]

\multirow{2}{*}{GF} & 46 $\pm$ 3 & 81 $\pm$ 4 & 100 $\pm$ 3 & 131 $\pm$ 4 & 137 $\pm$ 3 & 165 $\pm$ 5 \\ 
 & 0.0\% & 0.0\% & -1.0\% & -0.8\% & -0.7\% & -0.6\% \\[2ex]

\multirow{2}{*}{GF, const \To} & 44 $\pm$ 3 & 79 $\pm$ 4 & 93 $\pm$ 3 & 130 $\pm$ 4 & 138 $\pm$ 3 & 168 $\pm$ 6 \\ 
 & -4.3\% & -2.5\% & -7.9\% & -1.5\% & 0.0\% & 1.2\% \\ 
 \hline
\multicolumn{7}{p{\columnwidth}}{
Absolute values represent a ROI analysis and are given in ms (mean $\pm$ SD). Relative values for \Tt estimates represent the deviation to the reference.}
\end{tabular}
\end{table*}
confirm the visual impression. The relative error of the GF fit to the reference is $\SI{1}{\%}$ or less in all compartments. The mono-exponential fit, on the other hand, yields errors between $16$ and $\SI{28}{\%}$.

A practical limitation of the accurate GF fit is the necessity for additional \Bo and \To measurements. However, as has already been pointed out in \cite{petrovic,lebel}, the influence of \To on the GF result is relatively small. This finding is confirmed in Fig.~\ref{fig4} (right), where additional reconstructions were performed under the assumption of a constant \To value of $\SI{1000}{ms}$ for all pixels. Although the relative error in \Tt increases the stronger the chosen \To deviates from the true value, the results are surprisingly accurate and even for worst cases, the largest error in the \Tt map is still lower than the smallest error of the mono-exponential fit.

\subsection{Undersampling}
\begin{figure*}[t]
\centering
\ifmathvars
	\includegraphics[width=\textwidth]{fig5_sub}
\else
	\includegraphics[width=\textwidth]{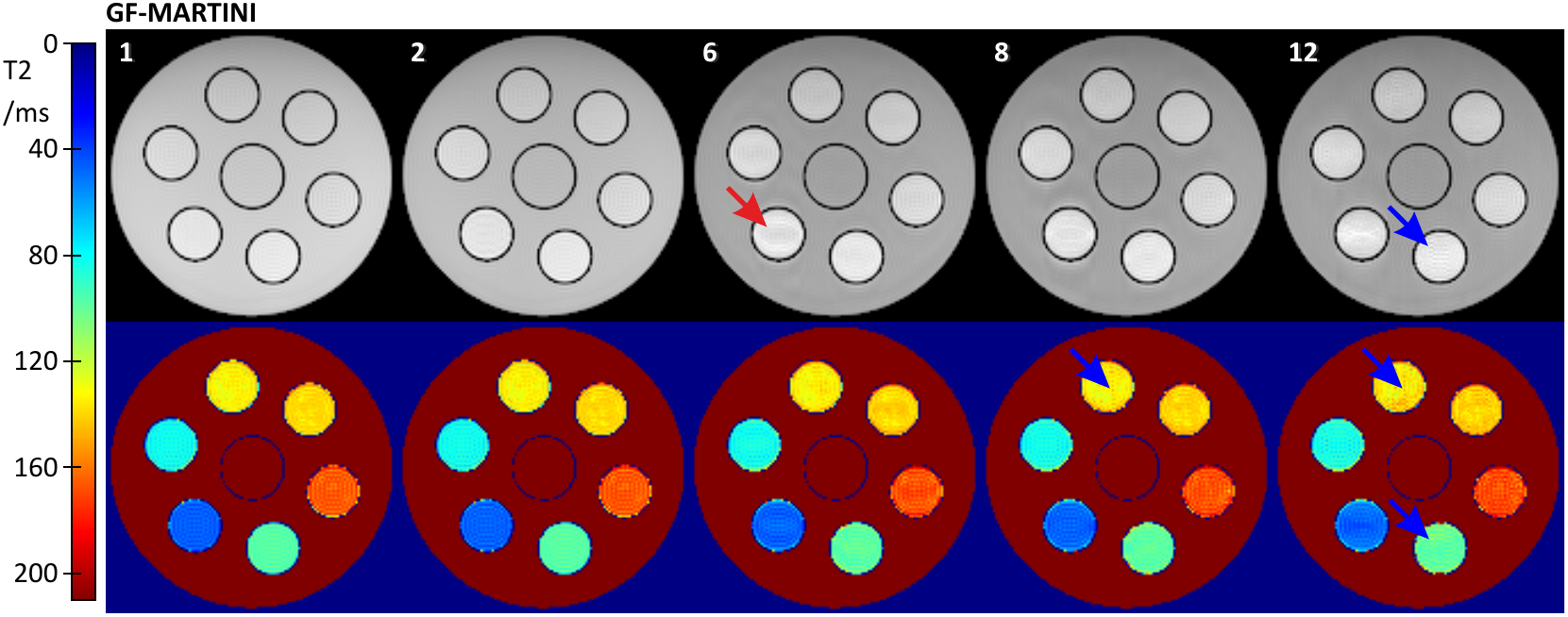}	
\fi
\caption{Spin-density and color-coded \Tt maps obtained by GF-MARTINI reconstructions (with validity mask) for undersampling factors of 1, 2, 6, 8 and 12. The corresponding measurement times were 12:54, 6:27, 2:09, 1:37, and 1:05\,min. Highly undersampled results reveal minor blurring (red arrow) as well as occasional ghosts (blue arrows). Mean \Tt values are accurate for all undersampling factors.}
\label{fig5}
\end{figure*}

For the relaxation phantom, Fig.~\ref{fig5}
demonstrates the results of the proposed GF-MARTINI method ($128$ frequency samples, validity mask, $3 \times 20$ CG iterations) for different degrees of undersampling. Again, effects of the limited oversampling in the frequency domain are most notable only in the spin-density map for the surrounding water compartment. However, this region was successfully identified as invalid after the first optimization block. Apart from that, all other regions are reconstructed with very high accuracy for all undersampling factors. Artifacts are barely notable except for some minor blurring around the compartment with the smallest \Tt (red arrow). For undersampling factor greater than $6$, some small vertical ghosts appear, as most notable in compartments $3$ and $6$ (blue arrows). For all cases a corresponding ROI analysis of \Tt values is given in Table~\ref{table2}.
\begin{table*}[!t]
\renewcommand{\arraystretch}{1.3}
\caption{\Tt relaxation times of a phantom using GF-MARTINI with different undersampling factors R}
\label{table2}
\centering
\begin{tabular}{lcccccc}
\hline
Compartment & 1 & 2 & 3 & 4 & 5 & 6 \\ 
\hline
R = 1 & 45 $\pm$ 3 & 81 $\pm$ 4 & 97 $\pm$ 3 & 132 $\pm$ 4 & 138 $\pm$ 3 & 166 $\pm$ 5 \\ 
R = 2 & 46 $\pm$ 4 & 82 $\pm$ 4 & 97 $\pm$ 4 & 132 $\pm$ 4 & 138 $\pm$ 3 & 166 $\pm$ 5 \\ 
R = 4 & 46 $\pm$ 4 & 82 $\pm$ 4 & 98 $\pm$ 4 & 133 $\pm$ 5 & 139 $\pm$ 3 & 168 $\pm$ 6 \\ 
R = 6 & 46 $\pm$ 5 & 82 $\pm$ 5 & 98 $\pm$ 4 & 132 $\pm$ 5 & 139 $\pm$ 4 & 167 $\pm$ 6 \\ 
R = 8 & 46 $\pm$ 4 & 82 $\pm$ 4 & 98 $\pm$ 4 & 132 $\pm$ 6 & 139 $\pm$ 4 & 167 $\pm$ 6 \\ 
R = 12 & 46 $\pm$ 4 & 82 $\pm$ 6 & 98 $\pm$ 7 & 133 $\pm$ 8 & 139 $\pm$ 5 & 168 $\pm$ 7 \\ 
 \hline
\multicolumn{7}{p{\columnwidth}}{
Values represent a ROI analysis and are given in ms (mean $\pm$ SD).}
\end{tabular}
\end{table*}
The resulting values are remarkably stable with deviations of less than $\pm\mbox{ }\SI{2}{ms}$ from the fully sampled SE reference for all compartments and undersampling factors.

\begin{figure*}[!t]
\centering
\ifmathvars
	\includegraphics[width=\textwidth]{fig6_sub}
\else
	\includegraphics[width=\textwidth]{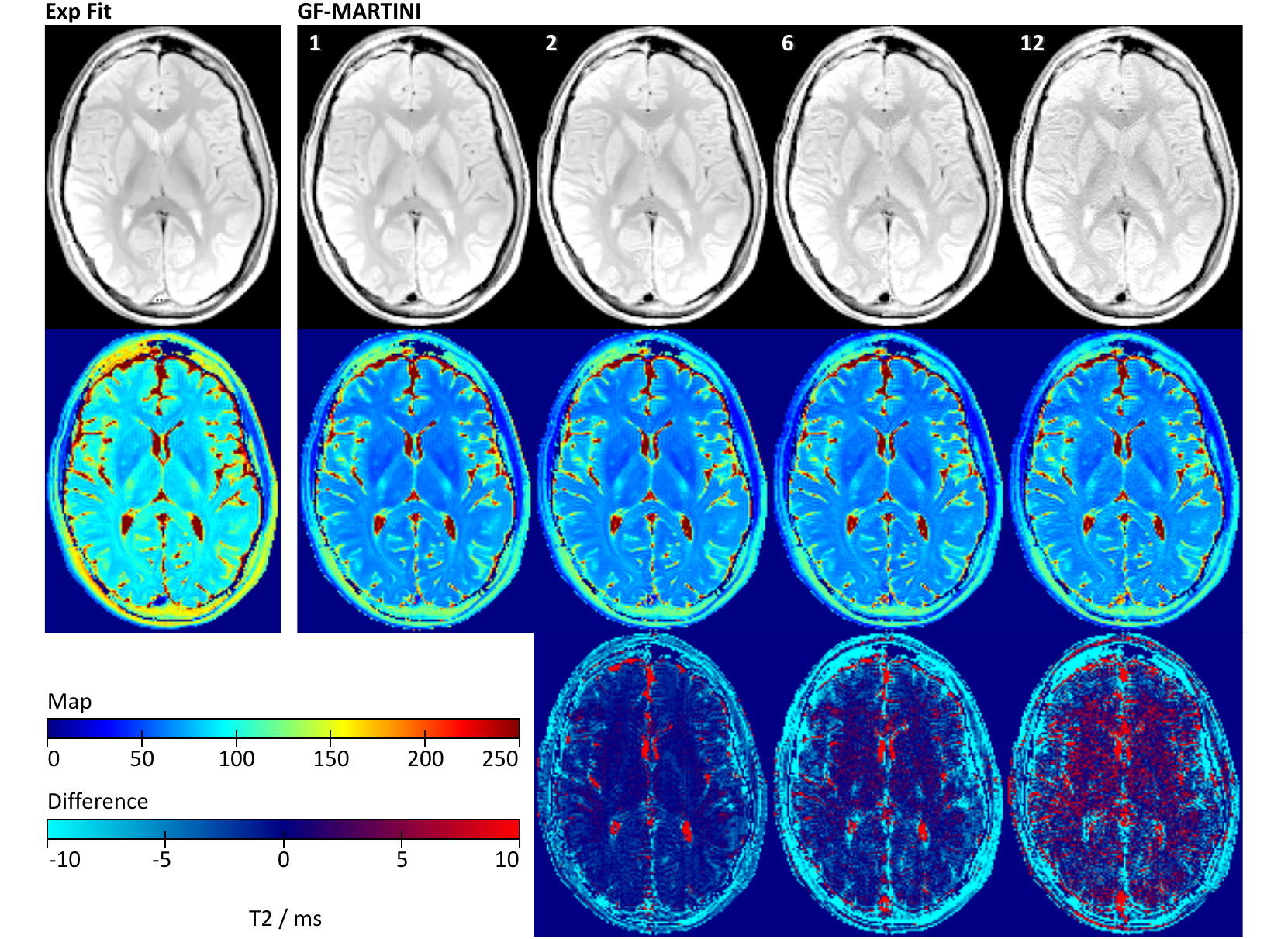}
\fi
\caption{(Top) Spin-density and (center) color-coded \Tt maps of the human brain obtained by (left) exponential fitting of fully sampled data and (right) GF-MARTINI reconstructions with validity mask for undersampling factors of 1, 2, 6 and 12. The corresponding measurement times were 12:54, 6:27, 2:09, and 1:05\,min. Reconstruction was performed with the use of a previously measured \To map, requiring an additional measurement time of 17:05\,min. (Bottom) \Tt difference maps with respect to the fully sampled reference.}
\label{fig6}
\end{figure*}

\begin{figure*}[!t]
\centering
\ifmathvars
	\includegraphics[width=\textwidth]{fig7_sub}
\else
	\includegraphics[width=\textwidth]{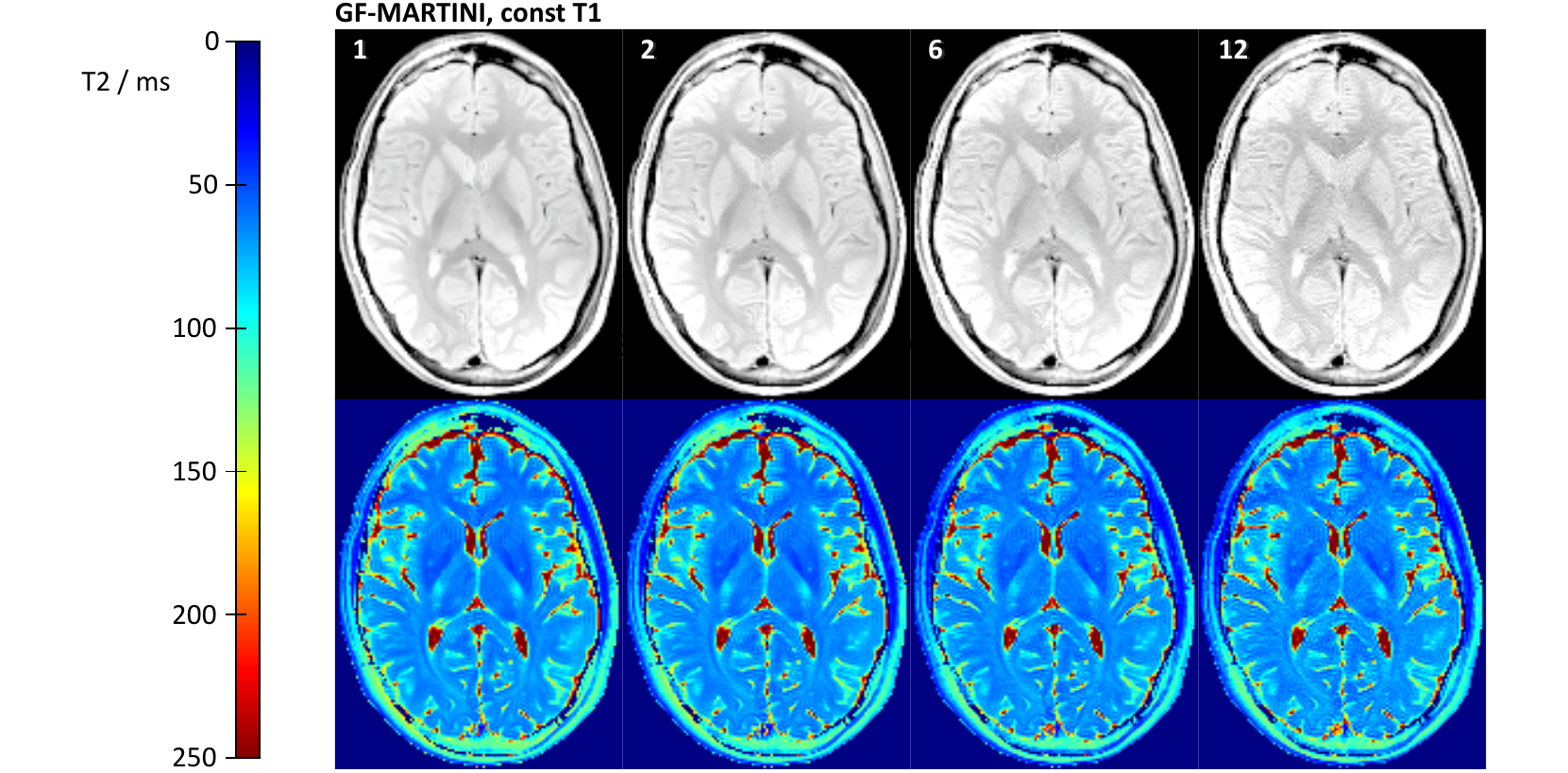}
\fi
\caption{Spin-density and color-coded \Tt maps of the human brain obtained by GF-MARTINI reconstructions with validity mask and constant \To = 1000\,ms for undersampling factors of 1, 2, 6 and 12. The corresponding measurement times were 12:54, 6:27, 2:09, and 1:05\,min.}
\label{fig7}
\end{figure*}

Corresponding results for a transverse section of the human brain are shown in Fig.~\ref{fig6}.
Whereas a pixel-wise exponential fit of the fully sampled MSE data (first echo removed) again leads to a systematic \Tt overestimation, the GF-MARTINI method presents with good accuracy for all undersampling factors. The most notable distortions are small vertical ghosts near the hemispheric fissure, which increase for higher undersampling factors. For the highest acceleration factor of $12$, also the spin-density map suffers from edge enhancement and blurring. A quantitative ROI analysis of \Tt values in various brain tissues is summarized in Table~\ref{table3}.
\newcommand*{\pb}[1]{\parbox{\cw}{\centering #1}}
\begin{table*}[!t]
\renewcommand{\arraystretch}{1.3}
\caption{\Tt relaxation times of the human brain using pixel-based methods and GF-MARTINI for different undersampling factors R.}
\label{table3}
\centering
\newlength{\cw}
\setlength{\cw}{1.2cm}
\begin{tabular}{lcccccc}
\hline
 &\pb{Anterior Cingulate}&\pb{Insular Cortex}&\pb{Thalamus}&\pb{Putamen}&\pb{Internal Capsule}&\pb{Frontal White Matter}\\
 \hline 
Mono-Exp & 101 $\pm$ 8 & 100 $\pm$ 10 & 84 $\pm$ 3 & 76 $\pm$ 5 & 96 $\pm$ 6 & 81 $\pm$ 3 \\[2ex]

GF & 73 $\pm$ 6 & 76 $\pm$ 8 & 60 $\pm$ 2 & 56 $\pm$ 4 & 70 $\pm$ 4 & 60 $\pm$ 2 \\[2ex]

\multicolumn{7}{l}{GF-MARTINI}\\
R = 1 & 73 $\pm$ 6 & 76 $\pm$ 8 & 59 $\pm$ 2 & 55 $\pm$ 4 & 69 $\pm$ 4 & 59 $\pm$ 2 \\ 
R = 2 & 71 $\pm$ 6 & 76 $\pm$ 8 & 59 $\pm$ 2 & 55 $\pm$ 4 & 69 $\pm$ 4 & 58 $\pm$ 3 \\ 
R = 6 & 71 $\pm$ 6 & 76 $\pm$ 8 & 60 $\pm$ 3 & 56 $\pm$ 4 & 69 $\pm$ 4 & 59 $\pm$ 3 \\ 
R = 12 & 71 $\pm$ 7 & 76 $\pm$ 8 & 60 $\pm$ 3 & 56 $\pm$ 5 & 70 $\pm$ 5 & 59 $\pm$ 3 \\[2ex]

\multicolumn{7}{l}{GF-MARTINI, const \To}\\
R = 1 & 73 $\pm$ 6 & 77 $\pm$ 8 & 59 $\pm$ 2 & 55 $\pm$ 4 & 69 $\pm$ 4 & 58 $\pm$ 2 \\ 
R = 2 & 72 $\pm$ 6 & 76 $\pm$ 8 & 59 $\pm$ 2 & 56 $\pm$ 5 & 68 $\pm$ 4 & 58 $\pm$ 2 \\ 
R = 6 & 72 $\pm$ 6 & 77 $\pm$ 9 & 60 $\pm$ 3 & 56 $\pm$ 5 & 69 $\pm$ 4 & 59 $\pm$ 3 \\ 
R = 12 & 72 $\pm$ 7 & 77 $\pm$ 8 & 60 $\pm$ 3 & 56 $\pm$ 5 & 69 $\pm$ 5 & 59 $\pm$ 3 \\ 
\hline
\multicolumn{7}{p{\columnwidth}}{
Values represent a ROI analysis and are given in ms (mean $\pm$ SD).}
\end{tabular}
\end{table*}
The mean \Tt values obtained by GF-MARTINI are in remarkably good agreement with the fully sampled reference and very stable for all undersampling factors. Moreover, Fig.~\ref{fig7}
depicts reconstructions of the same data with the assumption of a constant \To $= \SI{1000}{ms}$ for all pixels. It turns out that the GF-MARTINI reconstructions with constant \To are almost indistinguishable from the results obtained with a full \To map. The ROI analysis in Table~\ref{table3} confirms this observation as all measurable deviations between the two GF-MARTINI versions remain within $\pm\mbox{ } \SI{1}{ms}$.

\subsection{Reconstruction time}
The computation time of the algorithm mainly depends on the matrix size, number of receive coils and number of frequency samples in the \z -domain. In addition, the number of badly conditioned regions in the image can affect the reconstruction speed. Running on an Intel X5650 Xeon PC with $12$ virtual CPUs @ $\SI{2.67}{GHz}$, phantom and human brain MRI reconstructions in this study took between $2$ and $3\mbox{:}\SI{45}{min}$ when using $128$ frequency samples. Phantom calculations with $N_\omega= 512$ took up to $\SI{16}{min}$. However, a major speedup is expected when performing the calculations on graphic processing units due to the highly parallelizable structure of the algorithm.

\section{Discussion}
This work demonstrates the adaptation of the GF approach, which properly describes the signal decay of a multiple spin-echo MRI acquisition, for a model-based reconstruction from highly undersampled MSE data in order to achieve both fast and accurate \Tt mapping. For human brain MRI and in contrast to a mono-exponential model, the proposed GF-MARTINI method yields accurate \Tt values for undersampling factors of at least $6$, while reasonable \Tt maps may even be obtained for undersampling factors of $12$. 
Limitations for long \Tt relaxation times of fluids are caused by echo-time aliasing due to the finite sampling of the spin-echo train in the frequency domain. However, for \Tt values within a defined range of interest, the related errors can be avoided by moderate frequency-domain oversampling. The numerical instabilities caused by regions that strongly violate the GF model have further been shown to be controllable by adaptive masking and local dampening. Another strategy for a more tolerant performance for long \Tt relaxation times might result from apodization techniques as proposed in \cite{lukzen}. So far, however, preliminary trials let to a notable loss of \Tt accuracy.

In principle, the estimation of accurate \Tt values by GF-MARTINI requires prior knowledge of \To and \Bo field distributions. While deviations in \Bo indeed affect the \Tt decay of MSE experiments, the \To influence has been shown to be relatively small \cite{petrovic,lebel}. Depending on the application and actual range of expected \To values, the present findings support the practical feasibility of reliable reconstructions with the use of a single reasonable \To estimate. \Bo maps may be acquired at low resolution due to their smooth shape. It is also conceivable to treat the \Bo map as an additional unknown parameter of the cost function. By jointly estimating \Bo along with \Tt and spin-density, additional measurements may be avoided. However, while preliminary trials yielded promising results for pixel-based fits, we encountered stability issues when reconstructing from undersampled data.

A potential limitation of the proposed method is its dependency on properly chosen scaling and regularization variables that balance the gradient components of the nonlinear cost function. In the current implementation, these variables are kept at a fixed starting value, while a relatively large number of $3 \times 20$ CG iterations per column is used to counter small deviations from an optimal choice. For a more efficient and general application of the method, an automatic mechanism for adjusting the scaling would be highly desirable. Such development, however, is outside the scope of this work.

\ifieeestyle
	\appendices
	\appendix[Derivation of the cost function gradient]
\else
	\appendix
	\section{Appendix}
	\subsection{Derivation of the cost function gradient}
\fi
Optimization of the cost function $\Phi$ by means of the CG-DESCENT algorithm requires the implementation of the gradient of $\Phi$ with respect to the components in the scaled vector of real-valued unknowns $\mb{\tilde{x}}$. By combining all linear operations in \refeq{eq:cost} into a single system matrix $A_c$ for every coil element $c$, the cost function can be written as
\newcommand*{\tx}{\mb{\tilde{x}}}
\begin{equation}
\Phi (\tx) = \frac{1}{2}\sum_c \| A_c W(\tx) - \mb{s}_c \|_2^2
\end{equation}
The gradient of this term is given by
\begin{equation}
\nabla \Phi = \sum_c \Re{} \{ DW^H(\tx) A_c^H \left[ A_c W(\tx) -\mb{s}_c \right] \} 
\end{equation}
with $DW$ the Jacobian matrix of $W$ and $(\cdot)^H$ referring to the adjoint. With $(\overline{\cdot})$ denoting complex conjugation, the adjoint system matrix $A_c^H$ yields
\begin{equation}
A_c^H = F_{\omega}^H \overline{C}_c F_{xy}^H P
\end{equation}
where $F_{xy}^H$ and $F_{\omega}^H$ correspond to the inverse DFT in $x-y$ and frequency direction, respectively. Calculation of the Jacobian matrix $DW$ requires the partial derivatives of the scaled model function
\newcommand*{\rt}{\tilde{\rho}}
\newcommand*{\kt}{\tilde{k}_2}
\newcommand*{\G}{G_{\rt,\alpha,k_1,\kt}}
\begin{equation}
\G(z) = \frac{L_\rho \rt}{2} \left( 1 + \sqrt{\frac{\l1 \l2}{\l3 \l4}} \right)
\end{equation}
\vspace{-\baselineskip}
\begin{align}
\l1 &= 1 + z L_k \kt \\
\l2 &= 1 - z (k_1 + L_k \kt) \cos \alpha + z^2 k_1 L_k \kt\\
\l3 &= -1 + z L_k \kt\\
\l4 &= -1 + z(k_1 - L_k \kt) \cos \alpha + z^2 k_1 L_k \kt
\end{align}
which are given by
\begin{equation}
\frac{\partial}{\partial \rt} \G (z) = \frac{L_\rho}{2} \left( 1 + \sqrt{\frac{\l1 \l2}{\l3 \l4}} \right)
\end{equation}
and
\begin{equation}
\begin{split}
\frac{\partial}{\partial \kt} \G (z) = \frac{L_\rho \rt}{4} L_k
\left( 
\frac{\l2 z + \l1 \l5}{\l3 \l4}  \ldots \right. \\
\left. - z \frac{\l1 \l2}{\l3^2 \l4}
- \frac{\l1 \l2 \l5}{\l3 \l4^2}
\right)
\left(
\frac{\l1 \l2}{\l3 \l4}
\right)^{-1/2}
\end{split}
\end{equation}
\begin{equation}
\l5 = z^2 k_1 - z \cos \alpha
\end{equation}
for every pixel in image space.

\section*{Acknowledgment}
The authors thank Tom Hilbert from the EPFL Lausanne for valuable ideas to improve the computational efficiency of the reconstruction algorithm.

\ifieeestyle
	\bibliographystyle{IEEEtran}
\else
	\bibliographystyle{unsrt}
\fi
\bibliography{t2_map_gf_arXiv}

\end{document}